\documentclass[prl,twocolumn,superscriptaddress,floatfix,showpacs]{revtex4}

\usepackage{amsfonts,amsmath,amssymb,mathbbol}
\usepackage{latexsym}
\usepackage{graphicx}
\usepackage{bm,dcolumn}
\begin{document}

\title{Majorana modes in time-reversal invariant {\bf $s$}-wave
topological superconductors}

\author{Shusa Deng}
\affiliation{\mbox{Department of Physics and Astronomy, Dartmouth
College, 6127 Wilder Laboratory, Hanover, NH 03755, USA}}

\author{Lorenza Viola}
\affiliation{\mbox{Department of Physics and Astronomy, Dartmouth
College, 6127 Wilder Laboratory, Hanover, NH 03755, USA}}

\author{Gerardo Ortiz}
\affiliation{\mbox{Department of Physics, University of Indiana,
Bloomington, Indiana 47405, USA}}

\begin{abstract}
We present a time-reversal invariant $s$-wave superconductor
supporting Majorana edge modes.  The multi-band character of the model
together with spin-orbit coupling are key to realizing such a
topological superconductor.  We characterize the topological phase
diagram by using a partial Chern number sum, and show that the latter
is physically related to the parity of the fermion number of the
time-reversal invariant modes.  By taking the self-consistency
constraint on the $s$-wave pairing gap into account, we also establish
the possibility of a direct topological superconductor-to-topological
insulator quantum phase transition.
\end{abstract}


%74.78.-w 	Superconducting films and low-dimensional structures
%71.10.Pm 	Fermions in reduced dimensions (anyons, composite fermions, 
%                      Luttinger liquid, etc.) 
%73.20.At 	Surface states, band structure, electron density of states
%03.67.Lx  	Quantum computation architectures and implementations

\date{\today}
\maketitle

Since Majorana suggested the possibility for a fermion to coincide
with its own antiparticle back in 1937 \cite{Ettore}, the search for
the Majorana particle has catalized intense effort across particle and
condensed-matter physicists alike \cite{Frank}.  Particles either
constitute the building blocks of a fundamental physical theory or may
effectively emerge as the result of the interactions of a theory.  A
Majorana fermion is no exception to this principle, with neutrinos
potentially epitomizing the first view \cite{Avignone}, and localized
quasiparticle excitations in matter illustrating the second
\cite{Kanereview}.  Remarkably, Majorana fermions can give rise to the
emergence of non-Abelian braiding \cite{Moore}.  Thus, in addition to
their significance for fundamental quantum physics, interest in
realizing and controlling Majorana fermions has been fueled in recent
years by the prospect of implementing fault-tolerant topological
quantum computation \cite{Kitaev01,Nayak}.  As a result, a race is
underway to conclusively detect and characterize these elusive
particles.

A variety of condensed-matter systems hosting localized Majorana
elementary excitations have been proposed, notably certain quantum
Hall states \cite{Moore} and so-called {\em topological
superconductors} (TSs) \cite{Green,Ivanov}.  Unfortunately, these
exotic states of matter require the explicit breaking of time-reversal
(TR) symmetry and their physical realization seems to be at odds with
existent materials.  Such is the case, for instance, of
superconductors with $p_x+ip_y$ spin-triplet pairing symmetry. This
has not prevented researchers to pursue creative proposals that rely
on a combination of carefully crafted materials and devices.  Fu and
Kane \cite{Kane}, in particular, suggested the use of a (topologically
trivial) $s$-wave superconducting film on top of a three-dimensional
topological insulator (TI), which by proximity effect transforms the
non-trivial surface state of the TI into a localized Majorana excitation 
\cite{Kanereview,Sarma} (see also \cite{Volkov} for related early 
contributions).  While experimental realization of this idea awaits further 
progress in material science, alternative routes are being actively sought, 
including schemes based on metallic thin-film microstructures, quantum 
nanowires, and semiconductor quantum wells coupled to either a
ferromagnetic insulator, or to a magnetic field in materials with strong 
spin-orbit (SO) coupling \cite{Potter10etal}.

Our motivation in this work is to explore whether a path to TSs exists
based on conventional {\em bulk} $s$-wave spin-singlet pairing
superconductivity.  We answer this question by explicitly constructing
a model which, to the best of our knowledge, provides the first
example of a 2D TS with $s$-wave pairing symmetry, and supports
Majorana edge modes {\em without breaking TR symmetry} \cite{note1}.
The key physical insight is the {\em multi-band} character of the
model, in the same spirit of two-gap superconductors \cite{SMW}, but
with the SO coupling playing a crucial role in turning a topologically trivial two-gap superconductor into a non- trivial one. Our
results advance existing approaches in several ways.  First,
multi-band systems clearly expand the catalog of TI and TS materials.
Following the discovery of $s$-wave two-band superconductivity in 
MgB$_2$ in 2001, a number of two-gap superconductors ranging from
high-temperature cuprates to heavy-fermion and iron-based superconductors 
have already been characterized in the laboratory \cite{materials}, giving hope for 
a near-future material implementation.   Furthermore, from a theoretical standpoint,
our TR-invariant model also supports a direct TI-to-TS (first-order)
quantum phase transition (QPT), allowing one to probe these novel
topological phases and their surface states by suitably tuning control
parameters in the same physical system.

%%%%%%%%%%%%%%%%%%%%%%%%%%%%%%%%%%%%%%%%%%%%%%%

\emph{Exact solution with periodic boundary conditions.---} We
consider a TR-invariant two-band Hamiltonian of the form $H=H_{\sf
cd}+H_{\sf so}+H_{\sf sw}+ H.c.$, where
\begin{eqnarray}
H_{\sf cd}&=&\frac{1}{2}\sum_j(u_{cd} \psi_j^\dag \tau_x
\psi_j^{\;}-\mu \psi_j^\dag \psi_j^{\;}) -
t\hspace*{-0.5mm}\sum_{\langle i,j\rangle} \psi_i^\dag \tau_x
\psi_j^{\;}, \nonumber \\ H_{\sf so}&=&i \lambda
\hspace*{-1.5mm}\sum_{j,\nu=\hat{x},\hat{y}} \psi_j^\dag \tau_z
\sigma_\nu \psi_{j+\nu}, \nonumber \\ H_{\sf sw} &=& \sum_j
(\Delta_c c_{j,\uparrow}^\dag c_{j,\downarrow}^\dag+ \Delta_d
d_{j,\uparrow}^\dag d_{j,\downarrow}^\dag), 
\label{Ham}
\end{eqnarray}
represent the two-band ($c$ and $d$) dynamics, the SO interaction, and
$s$-wave superconducting fluctuations, respectively.  In the above
equations, $\mu$ is the chemical potential, $u_{cd}$  represents an onsite 
spin-independent ``hybridization term''  between the two bands, fermionic 
creation operators at lattice site $j$ (unit vectors $\hat{x},\hat{y}$) and
spin $\sigma=\uparrow, \downarrow$ are specified as
$c^\dagger_{j,\sigma}$ or $d^\dagger_{j,\sigma}$, depending on the
band, and ($\Delta_c$, $\Delta_d$) denote the mean-field $s$-wave
pairing gaps.  By letting $\psi_j \equiv
(c_{j,\uparrow},c_{j,\downarrow},d_{j,\uparrow},d_{j,\downarrow})^T$,
the Pauli matrices $\tau_{\nu}$ and $\sigma_{\nu}$ act on the orbital
and spin part, respectively.  Notice that we have implicitly assumed
that the intraband SO coupling strengths obey $\lambda_c = -\lambda_d
\equiv \lambda$.  In this way, in the limit where
$\mu=0=\Delta_c=\Delta_d$, $H$ reduces to a known model for a TI
\cite{Franz}.

For general parameter values and periodic
boundary conditions (PBC),  $H$ can be block-diagonalized by Fourier
transformation in both  $x$ and $y$.  That is, we can
rewrite $H=\frac{1}{2}\sum_{\bf{k}} (\hat{A}_{\bf{k}}^{\dag}
\hat{H}_{\bf{k}}\hat{A}_{\bf{k}}^{\;}-4\mu)$, with
$\hat{A}_{\bf{k}}^\dag=(c_{{\bf{k}},\uparrow}^\dag, c_{{\bf{k}},
\downarrow}^\dag,d_{{\bf{k}},\uparrow}^\dag,d_{{\bf{k}},
\downarrow}^\dag,c_{-{\bf{k}},\uparrow}^{\;},c_{-{\bf{k}},\downarrow}^{\;},
d_{-{\bf{k}},\uparrow}^{\;},d_{-{\bf{k}},\downarrow}^{\;})$, and
$\hat{H}_{\bf{k}}$ an $8 \times 8$ matrix.  An analytical
solution exists in the limit where the pairing gaps are $\pi$-shifted,
$\Delta_c=-\Delta_d\equiv \Delta$, since $\hat{H}_{\bf{k}}$ decouples
into two $4 \times 4$ matrices.  By introducing new canonical fermion
operators, $a_{{\bf{k}},\sigma}=\frac{1}{\sqrt{2}}(c_{{\bf{k}},\sigma}
+ d_{{\bf{k}},\sigma} )$, $b_{{\bf{k}},\sigma}=\frac{1}{\sqrt{2}}
(c_{{\bf{k}},\sigma} - d_{{\bf{k}},\sigma})$, we may rewrite
 $H=\frac{1}{2}\sum_{\bf{k}} (\hat{B}_{\bf{k}}^{\dag}
\hat{H}'_{\bf{k}} \hat{B}_{\bf{k}}^{\;}-4\mu)$, with
$\hat{B}_{\bf{k}}^\dag=(a_{{\bf{k}},\uparrow}^\dag, b_{{\bf{k}},
\downarrow}^\dag,a_{{-\bf{k}},\uparrow}^{\;},b_{{-\bf{k}},\downarrow}^{\;},
a_{{-\bf{k}},\downarrow}^\dag,
b_{{-\bf{k}},\uparrow}^\dag,a_{{\bf{k}},
\downarrow}^{\;},b_{{\bf{k}},\uparrow}^{\;})$, and
$\hat{H}'_{\bf{k}}=\hat{H}'_{1,\bf{k}} \oplus \hat{H}'_{2,\bf{k}}$,
with $\hat{H}'_{1,\bf{k}}$, $\hat{H}'_{2,\bf{k}} $ being TR of
one another, and
\begin{eqnarray*}
\hat{H}'_{1,\bf{k}}\hspace{-1mm}=\hspace{-1mm}\left (\begin{array}{cc}
\hspace{-1mm}m_{\bf{k}}\sigma_z-\mu+{\lambda_{\bf k}}\cdot
\vec{\sigma} & i\Delta\sigma_y\hspace{-1mm} \\
\hspace{-1mm}- i\Delta\sigma_y & -m_{\bf{k}}\sigma_z+\mu+{\lambda_{\bf
k}}\cdot \vec{\sigma}^*\hspace{-1mm}
\end{array} \right).
\label{4by4}
\end{eqnarray*}
Here, ${\lambda_{\bf k}}=-2\lambda(\sin{k_x}, \sin{k_y})$,
$m_{\bf{k}}=u_{cd}-2t(\cos{k_x}+\cos{k_y})$, and $\vec{\sigma} \equiv
(\sigma_x,\sigma_y)$.  The excitation spectrum obtained from
diagonalizing either $\hat{H}'_{1,\bf{k}}$ or $\hat{H}'_{2,\bf{k}}$ is
\begin{eqnarray}
\epsilon_{n,{\bf k}}=\pm \sqrt{m_{\bf{k}}^2\hspace{-0.7mm}+
\hspace{-0.7mm}\Omega^2\hspace{-0.7mm}+
\hspace{-0.7mm}|\lambda_{\bf{k}}|^2\hspace{-0.7mm}\pm2\sqrt{ m_{\bf{k}}^2
\Omega^2
\hspace{-0.7mm}+\hspace{-0.7mm}\mu^2|\lambda_{\bf{k}}|^2}},
\label{spectrum}
\end{eqnarray}
where the order $\epsilon_{1,{\bf k}} \leq \epsilon_{2,{\bf k}} \leq 0
\leq \epsilon_{3,{\bf k}} \leq \epsilon_{4,{\bf k}}$ is assumed and
$\Omega^2\equiv \mu^2 + \Delta^2$.  QPTs occur when the gap closes
($\epsilon_{2,{\bf k}}=0$, for general $\Delta \ne 0$), leading to the
critical lines determined by $m_{{\bf k}_c}=\pm \,\Omega$, where the
critical modes ${\bf k}_c \in \{(0,0),(0,\pi),(\pi,0),(\pi,\pi) \}$.
It is worth noticing that through a suitable unitary transformation
(see Eq.~(4) of Ref.~\cite{Sato}), the SO interaction in
Eq. (\ref{Ham}) is formally mapped into $p_x+ip_y$ and $p_x-ip_y$
intraband interaction, hinting at the existence of non-trivial
topological phases, as we demonstrate next.

\begin{figure}[ht]
\includegraphics[width=6.2cm]{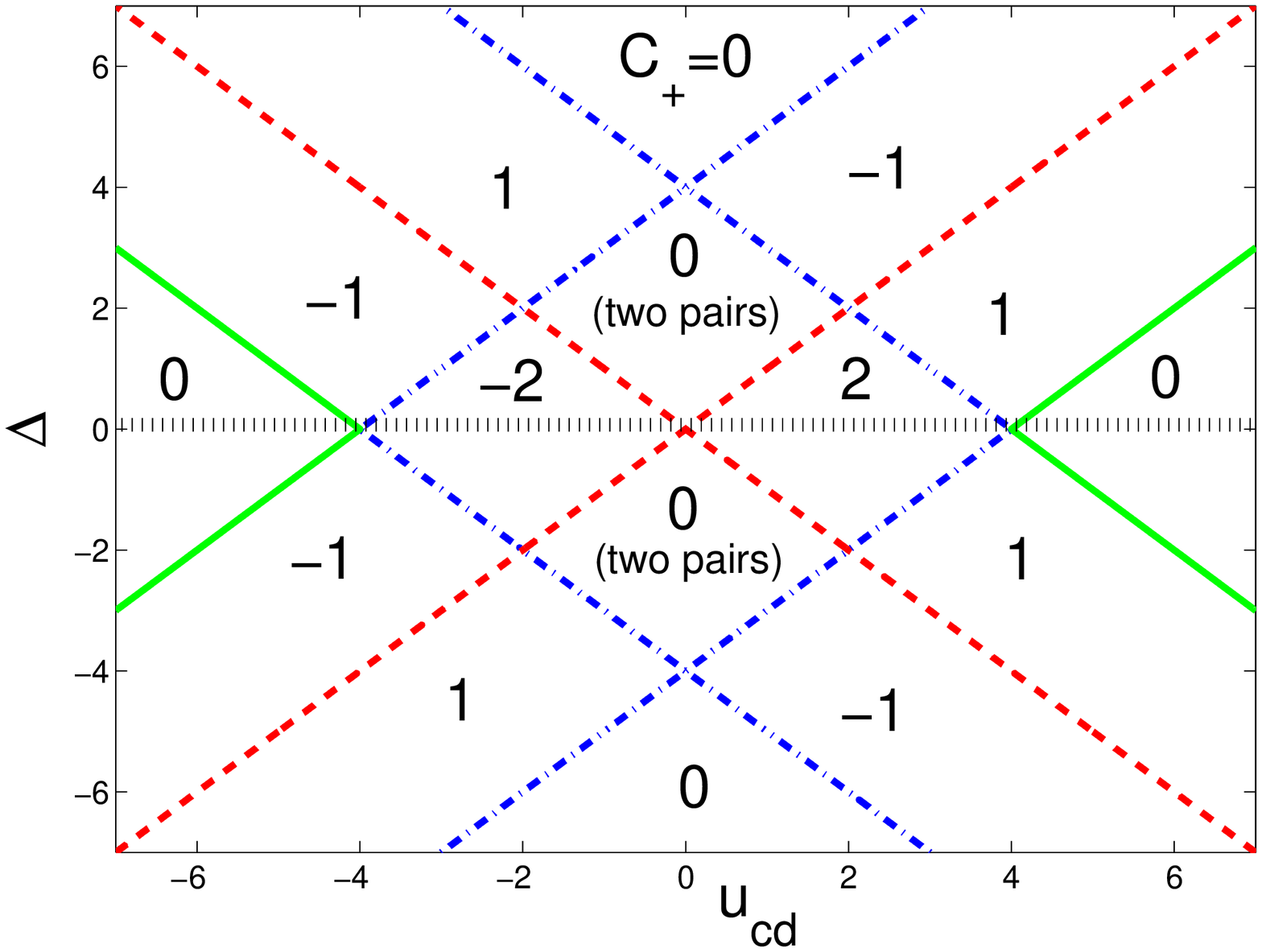}
\includegraphics[width=6.2cm]{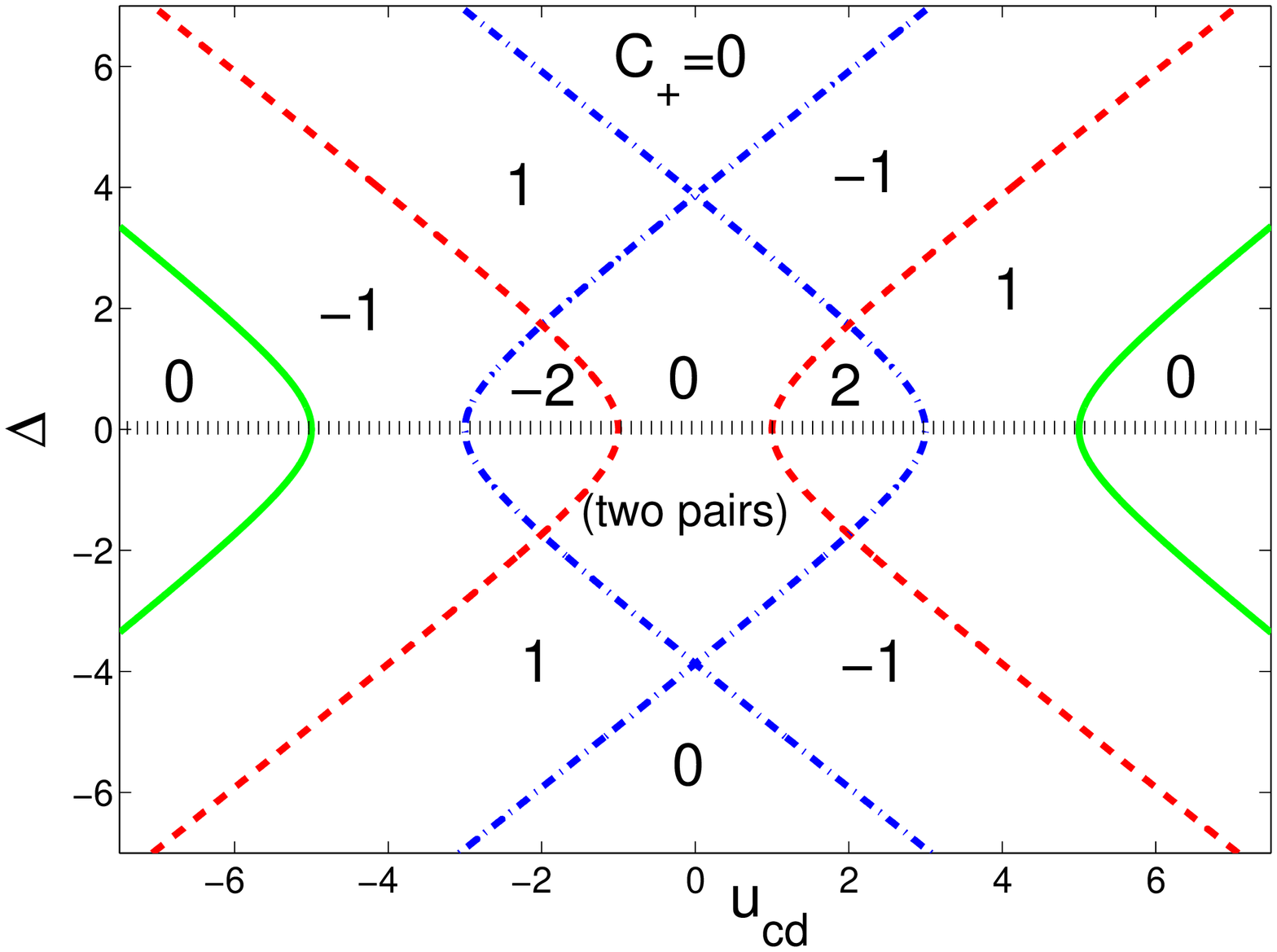}
\vspace*{-3mm} \caption{\label{u-1u0} (Color online) Topological
characterization of the phase structure of Hamiltonian $H$ via the
partial CN sum $C_+$ as a function of $u_{cd}$ and $\Delta$, with
$t=1$ and arbitrary $\lambda \ne 0$, for representative chemical
potentials $\mu=0$ (top) and $\mu=-1$ (bottom). The black (dashed)
line represents an insulator or metal phase, depending on the filling,
with $\Delta=0$.  CNs are calculated for $(N_x,N_y)=(100,100)$ lattice
sites. Note that we may have two pairs of edge modes with $C_+=0$.}
\end{figure}

\emph{Topological response.---} Since $H$ preserves TR invariance,
bands which form TR-pairs have opposite bulk Chern numbers (CNs)
$C_n$, leading to $\sum_{n \in {\sf occupied}} C_n=0$ (including both
$\hat{H}'_{1,\bf{k}}$ and $\hat{H}'_{2,\bf{k}}$).  Thus, introducing a
new $\mathbb{Z}_2$ topological invariant is necessary in order to
distinguish between trivial and TS phases.  In Ref. \cite{Roy}, the
parity of the sum of the {\em positive} CNs was considered, whereas in
Ref.~\cite{Kane06} an integral of the Berry curvature over {\em half}
the Brillouin zone for all the occupied bands was used.  Here, we
propose a different $\mathbb{Z}_2$ invariant which is
\emph{guaranteed} to work in the presence of TR: taking advantage of
the decoupled structure between TR-pairs, we use the CNs of the {\em
two occupied negative bands of $\hat{H}'_{1,\bf{k}}$ only} (say, $C_1$
and $C_2$) and define the following parity invariant:
\begin{equation}
\label{invariant}
P_C \equiv (-1)^{{\rm mod}_2(C_+)}, \;\;\; C_+ \equiv C_1+C_2.
\end{equation}
Let $|\psi_{n,{\bf k}}\rangle$ denote the band-$n$ eigenvector of
$\hat{H}'_{1,\bf{k}}$. Then the required CNs $C_n$, $n=1,2$, can be
computed as \cite{remark1}
\begin{eqnarray}
C_n=\frac{1}{\pi} \int_{-\pi}^\pi dk_x\int_{-\pi}^\pi dk_y \,
\text{Im} \,\langle \partial_{k_x}\psi_{n,{\bf k}}|\partial_{k_y}
\psi_{n,{\bf k}}\rangle.
\label{cn}
\end{eqnarray}
The resulting topological phase structure is shown in Fig.~\ref{u-1u0}
by treating the pairing gap $\Delta$ as a {\em free} control
parameter. In an actual physical system, $\Delta$ cannot be changed at
will, but only be found self-consistently by minimizing the free
energy (or ground-state energy at zero temperature). While we shall
return on this issue later, we first focus on understanding the
physical meaning of the above invariant and on establishing a
bulk-boundary correspondence for our model.

Interestingly, there is a direct connection between the invariant
$P_C$ and the fermion number parity of the TR-invariant modes. Without
loss of generality, let $\mu=0$, and focus on the ground-state fermion
number parity of the four TR-invariant points in the first Brillouin
zone, ${\bf k}_c$. Since $\hat{H}'_{1,\bf{k}}$ and
$\hat{H}'_{2,\bf{k}}$ are decoupled, we need only concentrate on the
ground-state parity property of $\hat{H}'_{1,\bf{k}}$. Let us
introduce the new basis: $\{ a_{{\bf
k}_c,\uparrow}^\dag |\text{vac}\rangle, b_{{\bf k}_c,\downarrow}^\dag
|\text{vac}\rangle, |\text{vac}\rangle, a_{{\bf k}_c,\uparrow}^\dag
b_{{\bf k}_c,\downarrow}^\dag |\text{vac}\rangle \}$. In this basis,
$\hat{H}'_{1,\bf{k}}$ becomes $\widehat{H}_{1,{\bf k}_c}=m_{{\bf
k}_c}\sigma_z \oplus \Delta \sigma_x$, with eigenvalues $\pm m_{{\bf
k}_c}, \pm\Delta$, and an identical matrix for $\hat{H}'_{2,{\bf
k}_c}$ in the TR-basis $\{ -a_{{\bf k}_c,\downarrow}^\dag |\text{vac}\rangle,b_{{\bf k}_c,\uparrow}^\dag|\text{vac}\rangle, 
|\text{vac}\rangle, -a_{{\bf k}_c,\downarrow}^\dag b_{{\bf k}_c,\uparrow}^\dag |\text{vac}\rangle \}$.  When $|m_{{\bf k}_c}| > |\Delta|$, the ground
state of each mode ${\bf k}_c$ is in the sector with odd fermion
parity, $P_{{\bf k}_c}=e^{i \pi (a_{{\bf k}_c,\uparrow}^\dag a_{{\bf
k}_c,\uparrow}^{\;}+b_{{\bf k}_c,\downarrow}^\dag b_{{\bf
k}_c,\downarrow}^{\;})}=-1$, otherwise it is in the sector with even
fermion parity, $P_{{\bf k}_c}=1$.  By analyzing the relation between
$|m_{{\bf k}_c}|$ and $|\Delta|$ for each ${\bf k}_c$, we can see that
the TS (trivial) phases with $P_C=-1(1)$ correspond to the ground state
with $\prod_{{\bf k}_c} P_{{\bf k}_c} \equiv P_F=-1(1)$.  Thus, our
$\mathbb{Z}_2$ invariant coincides with the fermion number parity of
the four TR-invariant modes {\em from one representative of each
Kramer's pairs}, consistent with the fact that only a partial CN sum
can detect TS phases in the presence of TR symmetry.  While the
relation between non-trivial topological signatures (such as the
fractional Josephson effect) and the local fermion parity of Majorana
edge states has been discussed in the
literature~\cite{Kitaev01,Kane09,Lee11}, invoking the fermion number
parity of the TR-invariant modes in {\em bulk periodic systems} to
characterize TS phases has not, to the best of our knowledge.

%%%%%%%%%%%%%%%%%%%%%%%%%%%%%%%%%%%%%%%%%%%%

\begin{figure}[t]
\includegraphics[width=9.5cm]{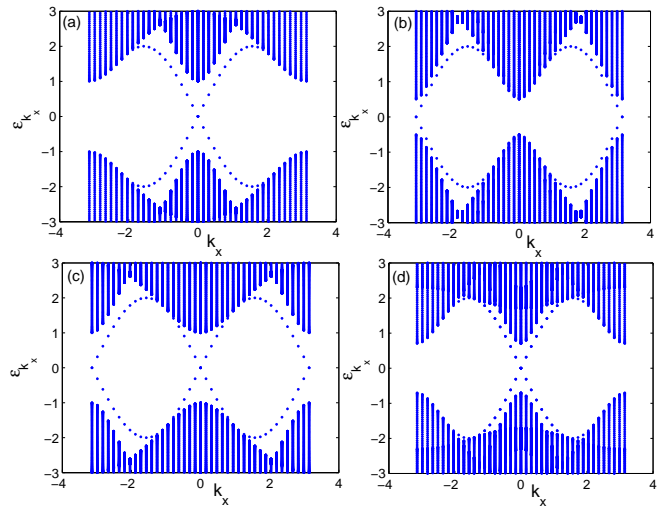}
\vspace*{-5mm}\caption{\label{tso} (Color online) Excitation spectrum
of Hamiltonian $H$ on a cylinder for $\mu=0, t=1, \lambda=1$.  Panel
(a), $C_+=1$: $\Delta=2$, $u_{cd}= 3$; Panel (b), $C_+=-1$:
$\Delta=2.5$, $u_{cd}=2$; Panel (c), $C_+=0$: $\Delta=2$, $u_{cd}= 1$;
Panel (d), $C_+=2$: $\Delta=0.8$, $u_{cd}=1.5$. Note that the bulk gap
scales as $\min(\lambda,\Delta)$. The number of lattice sites $(N_x,
N_y)=(40,100)$. }
\end{figure}

\emph{Open boundary conditions and edge states.---} A hallmark of a TS
is the presence of an {\em odd} number of pairs of gapless helical
edge states, satisfying Majorana fermion statistics. Thus, in order to
understand the relation between $P_C$ (or $P_F$) and the parity of the
number of edge states, {\em i.e.}, a bulk-boundary correspondence, we
study the Hamiltonian $H$ on a cylinder.  That is, we retain PBC only
along $x$, and correspondingly obtain the excitation spectrum,
$\epsilon_{n,k_x}$, by applying a Fourier transformation in the
$x$-direction only.  For simplicity, let us again focus on the case
$\mu=0$. The resulting excitation spectrum is depicted in
Fig.~\ref{tso} for representative parameter choices.  Specifically,
for {\em odd} $P_C$ ($C_+=1$ in panel (a) and $C_+=-1$ in panel (b),
respectively), $H$ supports one TR-pair of helical edge states on each
boundary, corresponding to the Dirac points $k_x=0$ (a) and $k_x=\pi$
(b). Different possibilities arise for {\em even} $P_C$.  While
$C_+=0$ can clearly also indicate the absence of edge states, in panel
(c) one TR-pair of helical edge states exists on each boundary for
both Dirac points $k_x=0,\pi$ (for a total of two pairs, as also
explicitly indicated in Fig.~\ref{u-1u0}).  In panel (d) ($C_+=2$),
both TR-pairs of helical edge states correspond to the Dirac point
$k_x=0$ instead.  Since, as remarked, our Hamiltonian exhibits
particle-hole symmetry, the equation
$\gamma_{\epsilon_{n,{k}_x}}^{\;}=\gamma^\dag_{-\epsilon_{n,{k}_x}}$
holds for each eigenvalue $\epsilon_{n,{k}_x}$, where
$\gamma_{\epsilon_{n,{k}_x}}^{\;}$ is the associated quasi-particle
annihilation operator.  Thus, for zero-energy edge states,
$\gamma_{0}^{\;}=\gamma^\dag_{0}$, indicating that the edge states in
our system satisfy Majorana fermion statistics.

%%%%%%%%%%%%%%%%%%%%%%%%%%%%%%%%%%%%%%%%%%%%

\begin{figure}[t]
\includegraphics[width=8cm]{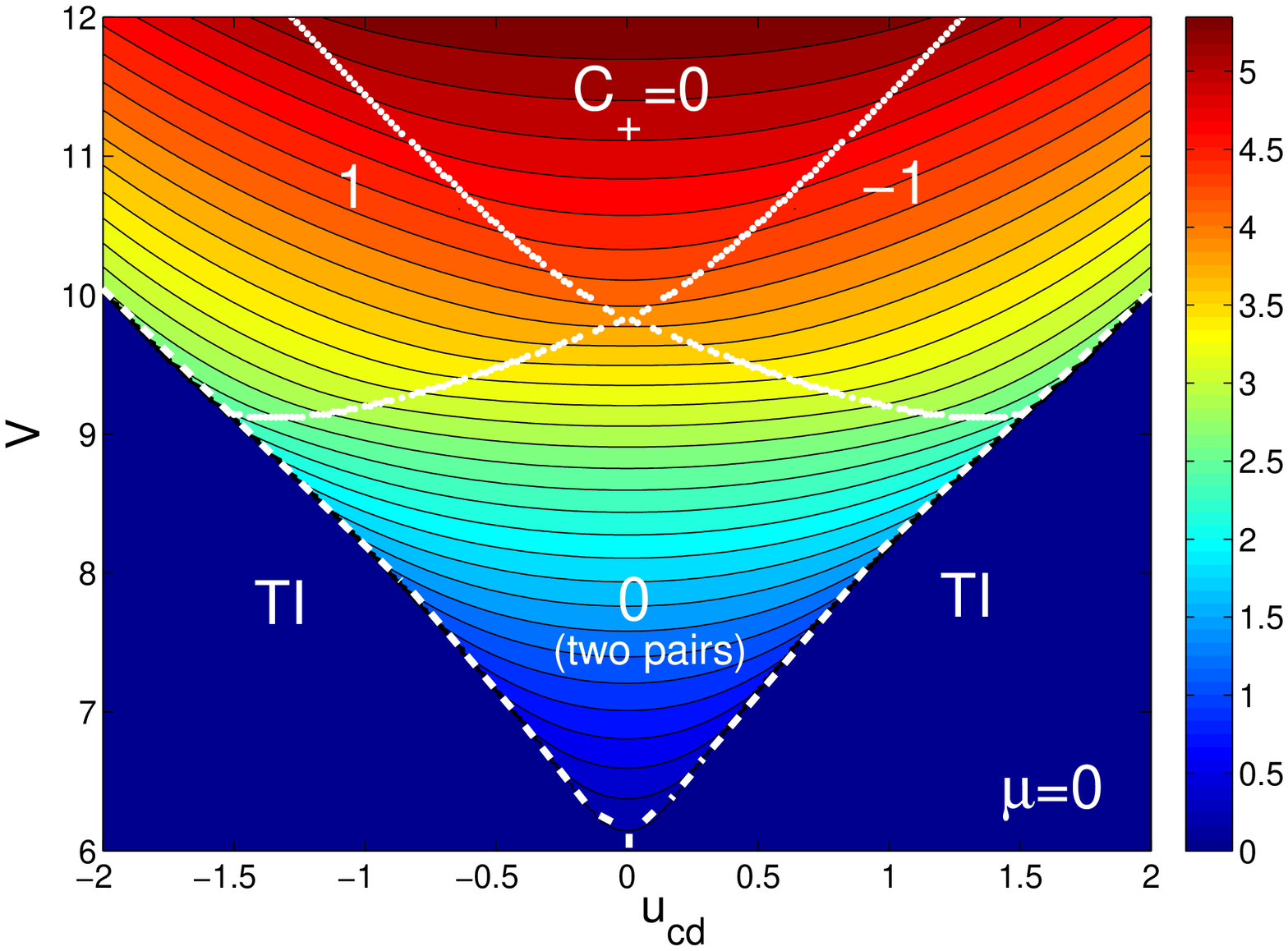}
\includegraphics[width=8cm]{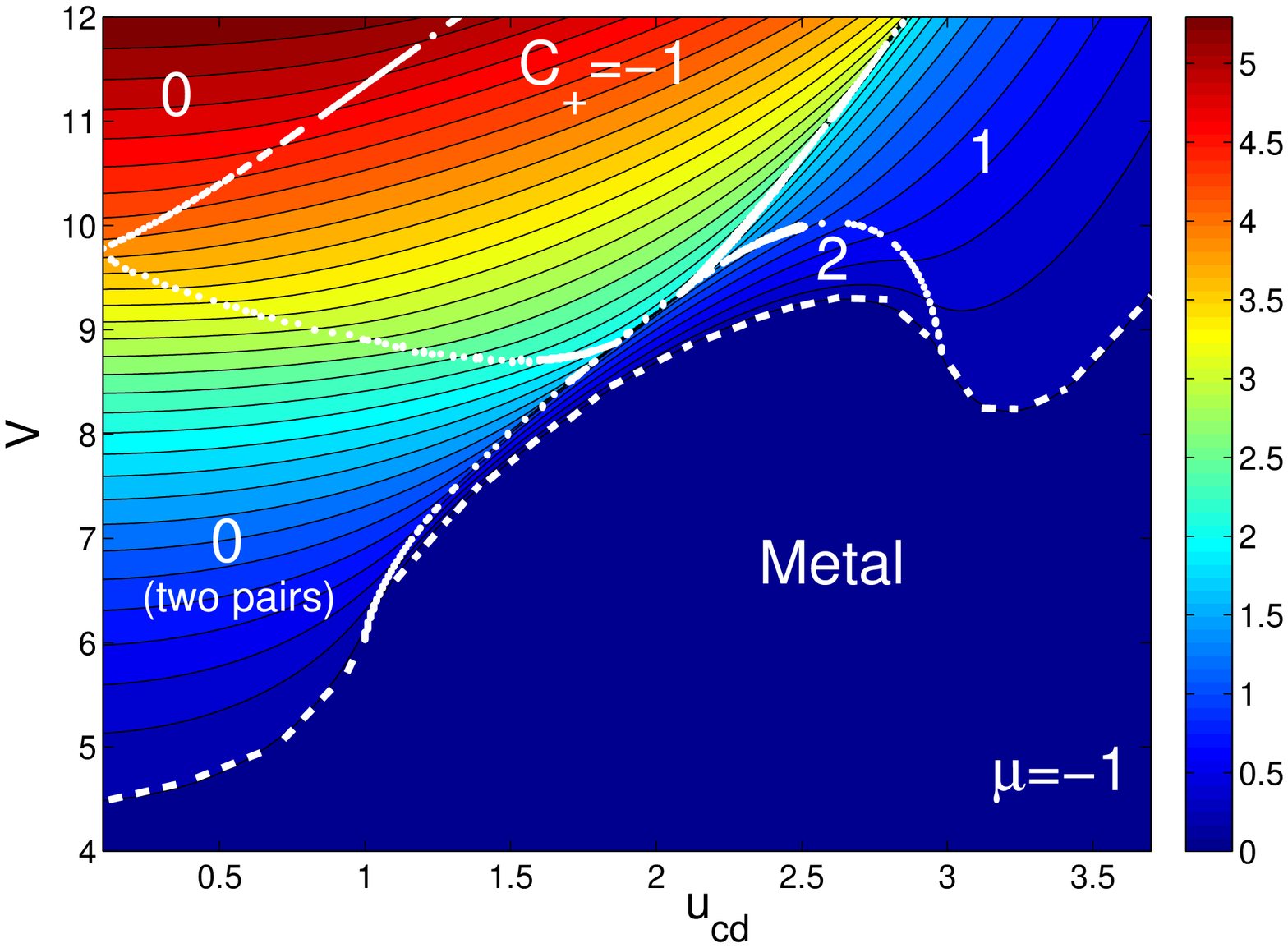}
\vspace*{-5mm} \caption{\label{scu-1} (Color online) Phase diagram as a function of
$u_{cd}$ and $V$ with the pairing gap $\Delta$ calculated
self-consistently.  The magnitude of $\Delta$ is represented by a
color, whose scale is indicated on the side.  The number of lattice
sites $(N_x, N_y)=(80,80)$.}
\end{figure}

\emph{Phase diagram with self-consistent pairing gap.---} Within BCS
mean-field theory, let $V\equiv V_{\bf{k},\bf{k'}} >0$ denote the
effective attraction strength in each band.  Then the pairing gap
$\Delta = \Delta_c = - V \langle c_{\bf{k},\uparrow}
c_{\bf{-k},\downarrow} \rangle = - \Delta_d$, and the ground-state
energy can be written as
%\begin{eqnarray*}
$E_g= 2 N_x N_y ({\Delta^2}/{V})+\sum_{{\bf k}} (\epsilon_{1,{\bf k}}
+ \epsilon_{2,{\bf k}}-2\mu).$
%\end{eqnarray*}
The first (constant) term is the condensation energy, which was
neglected in $H$.  By using Eq. (\ref{spectrum}) and minimizing $E_g$,
we obtain the stable self-consistent pairing gap $\Delta$ as a
function of the remaining control parameters \cite{Kub}. The resulting
zero-temperature phase diagram is shown in Fig.~\ref{scu-1}.  For
$\mu=0$ (top panel), the average fermion number is consistent with
half-filling, and thus with an insulating phase when $\Delta=0$. In
particular, when $0< |u_{cd}| < 4$, the ground state is known to
correspond to a TI phase \cite{Franz}.  Interestingly, without
self-consistency, the TI {\em cannot} be turned into a TS directly, as
shown in the top panel of Fig.~\ref{u-1u0}.  However, after
self-consistency is taken into account, the topologically trivial
phase with $C_+=\pm 2$ disappears, and a first-order QPT can connect
the two phases.  For $\mu=-1$ (bottom panel), the average fermion
number is found to be less than half-filling, realizing a metallic
phase when $\Delta=0$. Derivatives of the ground-state energy indicate
that {\em all} QPTs, except the TI-to-TS phase transition, are
continuous.

\emph{Discussion.---} A number of remarks are in order.
First, while the choice of SO coupling strengths and $s$-wave pairing
gaps obeying $\lambda_c = -\lambda_d$ and $\Delta_c = -\Delta_d$
affords a fully analytical treatment, relaxing these conditions 
may be necessary to make contact with real
materials.  Numerical results on a cylinder show that the level
crossing of the Majorana edge states in the TS phase is {\em robust}
against perturbations around $\lambda_c = -\lambda_d$, including the
possibility that the SO coupling vanishes in one of the bands. TS
behavior also persists if $|\Delta_c|- |\Delta_d|\ne0$, as long as the
phase difference between pairing gaps is $\pi$. In the presence of a 
phase mismatch $\varepsilon$, edge modes are found to become
gapped, with a minimal gap that scales linearly with $\varepsilon$.  
Interestingly, however, preliminary results indicate that adding a suitable 
Zeeman field can allow (at the expense of breaking TR invariance) 
gapless Majorana excitations to be restored, with a precise tuning of the 
phase difference being no longer required.  It is also worth noting that 
one can reinterpret the band index in $H$ as a layer index, and so $H$ 
may be thought of as describing a {\em bilayer} of superconductors with 
phase-shifted pairing gaps, and an interlayer coupling $H_{\sf cd}$.  
Beside establishing a {\em formal} similarity with the scenario discussed 
by Fu and Kane \cite{Kane}, such an interpretation may offer additional 
implementation flexibility, as the possibility to control the superconducting
and SO couplings by an applied gate voltage has been 
demonstrated recently \cite{Dagan}. 

Second, we have thus far restricted to 2D systems in order to simplify
calculations.  Preliminary results indicate that a qualitatively 
similar behavior (that is, the possibility of even/odd numbers of pairs 
of gapless Majorana surface states) also exists for 3D systems obtained from 
a natural extension of our 2D Hamiltonian.  It is especially suggestive to note 
that a $\pi$ phase shift in the order parameter across two bands is also believed 
to play a key role in iron pnictide superconductors \cite{Mazin}, hinting at possible 
relationships between TS behavior and so-called {\em $s_\pm$ pairing symmetry}.
While a more detailed investigation is underway, it is our hope that multi-band 
superconductivity may point to new experimentally viable venues for exploring 
topological phases and their exotic excitations.

%%%%%%%%%%%%%%%%%%%%%%%%%%%%%%%%%%%%%%%%%%%%%%

It is a pleasure to thank Charlie Kane for insightful discussions.
Support from the NSF through Grants No. PHY-0903727 (to L. V.) and 1066293 (Aspen Center for Physics) is
gratefully acknowledged.
 
%%%%%%%%%%%%%%%%%%%%%%%%%%%%%%%%%%%%%%%%%%%%%%%

{}

\end{document}